\documentclass[aps,prb,superscriptaddress,showpacs,floatfix]{revtex4-2}
\usepackage{amsbsy,amssymb,amsmath,bm}
\usepackage{float}
\usepackage{epsf}
\usepackage{graphicx,color}
\usepackage[caption = false]{subfig}
\usepackage{amsfonts}
\usepackage{amssymb}
\usepackage{amsmath}
\usepackage[T1]{fontenc}
\usepackage{xspace}
\usepackage{epsfig}
\usepackage{ctable}
\usepackage{enumitem}

\begin{document}

\title{Unified intermediate coupling description of the pseudogap and the
strange metal phases of cuprates}
\author{H. C. Kao}
\email{hckao@phy.ntnu.edu.tw}
\author{Dingping Li }
\email{lidp@pku.edu.cn}
\author{Baruch Rosenstein }
\email{baruchro@hotmail.com}
\date{\today }

\begin{abstract}
A one band Hubbard model with intermediate coupling is shown to describe the
two most important unusual features of a normal state: linear resistivity
strange metal and the pseudogap. Both the spectroscopic and transport
properties of the cuprates are considered on the same footing by employing a
relatively simple postgaussian approximation valid for the intermediate
couplings $U/t=1.5-4$ in relevant temperatures $T>100{\rm K}.$ In the doping range
$\ p=0.1-0.3$, the value of $U$ is smaller than that in the parent material.
For a smaller doping, especially in the Mott insulator phase, the coupling
is large compared to the effective tight binding scale and a different
method is required. This scenario provides an alternative to the paradigm
that the coupling should be strong, say $U/t>6$, in order to describe the
strange metal. We argue that to obtain phenomenologically acceptable
underdoped normal state characteristics like $T^{\ast }$, pseudogap values,
and spectral weight distribution, a large value of $U$ is detrimental.
Surprisingly the resistivity in the above temperature range is linear $\rho
=\rho _{0}+\alpha \frac{m^{\ast }}{e^{2}n\hbar }T$ with the "Planckian"
coefficient $\alpha $ of order one.
\end{abstract}

\maketitle

\affiliation{\textit{Physics Department, National Taiwan Normal University,
Taipei 11677, Taiwan, R. O. C}}

\affiliation{\textit{School of Physics, Peking University, Beijing 100871,
China}}
\affiliation{\textit{Collaborative Innovation Center of Quantum
Matter, Beijing, China}}

\affiliation{\textit{Electrophysics Department, National Chiao Tung
University, Hsinchu 30050, Taiwan, R. O. C}}


Physical nature of the d-wave pairing in high $T_{c}$ cuprates remains a
hotly debated topic in condensed matter physics. However it was noticed
early on that the normal state is as unusual as the superconducting one. The
two main unusual features, broadly referred to as the pseudogap\cite%
{pseudogaprev} and the strange metal \cite{strangerev}, cannot be described
in the customary framework of the Landau liquid. A pseudogap appears near
the antinodal points below temperature $T^{\ast }$ and increases in the
underdoped regime $p<p_{opt}$ ($p_{opt}=0.16$) towards the Mott insulator
phase reaching values $\Delta =100-200{\rm meV}$ \cite{Tallon,Shen14}.
A natural explanation relies on the short range antiferromagnetic (AF)
order, since the pseudogap regime borders the Mott insulator phase at very
low doping. It was observed\cite{Lifshitz} that the Fermi surface fractures
into small "pockets" for doping below the Lifshitz point $p^{\ast }$%
precisely when the pseudogap vanishes. This is described quite well by
variants of the phenomenological RVB model\cite{YRZ} derived from a
Hubbard model with on site repulsion $U$ (and various hopping parameters $%
t,t^{\prime }$...) within a generalization of the mean field theory largely
preserving the quasiparticle picture\cite{Metzner07}.

The hallmark of the strange metal is linear resistivity in the $100-450{\rm K}$
temperature range at intermediate and relatively large doping, generally
above $T^{\ast }$. It is believed that the quasiparticle picture should be
properly modified in this phase. Approaches, like the marginal Fermi liquid%
\cite{Varma}, quantum criticality\cite{QCP}, and Planckian dissipation\cite%
{Planck}, explain part of the experimental results, but typically start from
Hamiltonians different from those used to describe the pseudogap physics.
Thus each particular aspect of the normal state can be captured by a
particular phenomenological model to provide a consistent description of the
whole normal state (not including superconductivity) phase diagram, see Fig.
sm2 in Supplemental Material I (SMI), but a single theory is still a
challenge. The main problem seems to be the conflict between two "paradigms". It is widely accepted that \textit{strong} electron
correlations (in our case on-site repulsion $U$ larger than the
band width of $\sim 5t$) play a central role in both the spin fluctuation
theory of the d-wave superconductivity and in the normal state properties.
It seems consistent with the measured values of $U\sim 1-3{\rm eV}$ for the parent
material ($x=0$) Mott gap in many cuprates. First principle derivations
(almost exclusively at zero doping) of the "mesoscopic" one band Hamiltonian
support a strong coupling\cite{Nilsson19}. For one layer cuprates, one
obtains $U=7.7t$ for $La_{2}CuO_{4}$ ($T_{c}=34{\rm K}$ at $p_{opt}$), $U=7.2t$
for $HgBa_{2}CuO_{4}$ ($T_{c}=96{\rm K}$). The trend upon the inclusion of the
long range Coulomb interactions on the microscopic level however is towards
lower values\cite{Jang16}: $U=6.6t$ for $La_{2}CuO_{4}$, $U=4.5t$ for $%
HgBa_{2}CuO_{4}$. Electron doped and infinite layer cuprates have
significantly lower values $U/t=1.3-3$, so that in some cases the Mott
insulator phase is missing\cite{eldop}, e.g. $Nd_{2}CuO_{4}$ ($%
T_{c}=24{\rm K})$ one obtains~\cite{Jang16} $U/t=2.6$.

Upon doping, the effective coupling strength $U$ in the mesoscopic level is
expected to decrease. Recently a first principle study of doped cuprates $%
La_{2-p}Sr_{p}CuO_{4}$, $p=0.25$, was performed\cite{MarkiewiczDFT18}.
Although the values of $U$ were not explicitly calculated, reduction of the
gap at crystallographic $X$ point by a factor of $2.5$ compared to the
parent material indicates a lower value. To graphically demonstrate the
qualitative distinction between the strongly and intermediate coupling
regimes, let's look at the crossover temperature $T^{\ast }$ in the simplest
Hubbard model at half filling. Interpolating between the early "simplistic"
Hartree-Fock (HF) approximation at small coupling and the nonlinear $%
\sigma $ model at large coupling \cite{Dupuis}, one expects a maximum to
appear at $U=5t$, (blue curve in Fig.1). The gaussian perturbation
theory\cite{sym} at intermediate coupling interpolated with the CDMFT
diagrammatic approach\cite{Tremblay19} at strong coupling gives rise to
correction (purple curve in Fig.1). The $U$ dependence of $T^{\ast}$
obtained from these two calculations share the same feature: rising
monotonically at intermediate coupling and decreasing at strong coupling.
This separates the (Slater) weakly correlated from the (Mott-Heisenberg)
strongly correlated domains. It is sometimes referred to\cite%
{MarkiewiczSR17,Jang16} as the "Mott-Slater transition". The feature
remains intact (with typically lower values of $T^{\ast })$ at nonzero
doping.

The high values of the coupling at optimal doping are naturally favoured in
the Hubbard model description of both the normal and superconducting states.
Recently, however, serious doubts were cast on this possibility. Employing
the tensor network\cite{tensor}, constrained path Quantum Monte Carlo (CPQMC), density matrix renormalization group (DMRG)\cite{MC}  and the
strong coupling diagram technique (SCDT)\cite{strong} methods, it was demonstrated that the d-wave
superconductivity is superseded by other phases at least for $U>6t$. In the
Slater regime the Hubbard model does exhibit d-wave superconductivity
within perturbation theory \cite{Kivelson}, but its $T_{c}$ is too low.
Whether the Hubbard model supports sufficiently strong d-wave
superconductivity at intermediate region $2<U/t<5$ is still an open question
considering recent opposite claims\cite{Kozik21,phonon}. In addition the
coupling strength is poorly correlated with the observed values in cuprates,
e.g. the highest $T_{c}=133K$ (under ambient conditions) tri-layer
superconductor\cite{Jang16} $HgBa_{2}Ca_{2}Cu_{3}O_{8}$ has DFT estimated
(zero doping) value of $U/t=2.5$ only. To quote the authors, "Our results
suggest that the strong correlation enough to induce Mott gap may not be a
prerequisite for the high-$T_{c}$ superconductivity." Recent alternatives
include the apical phonon \cite{phonon} and polaron \cite{polaron}
mechanisms.

A general feature of the strong coupling scenario is that the spectrum is
expected to be greatly reconstructed and might not contain well defined
quasiparticles. However quasiparticles are observed in numerous experiments
perhaps excluding the strange metal regions\cite{nonquasiparticle}. The
ARPES experiments and transport properties are "phenomenologically"
described by the quasiparticle picture thus tacitly assuming a weak
coupling. This includes description of small Fermi "pockets" in the
underdoped regime, and the Landau liquid in highly overdoped samples\cite%
{YRZ}. In addition the order of magnitude of the pseudogap (up to $200{\rm meV}$
at $p=0.05$) is smaller than $U$ for strong coupling, thus favoring an
intermediate coupling option $U/t\sim 2-4$. In this case the quasiparticles
are well defined and symmetrized mean field approach\cite{sym} may be used
to describe the temperature range above $100{\rm K}$ and doping above $p=0.1$.
This includes physics of the strange metal and pseudogap for
sufficiently large Fermi arcs (pockets).

In this letter the normal state properties of a generic (one layer) hole
cuprate (e.g. $HgBa_{2}CuO_{4+p }$) in the doping ($0.1<p<0.3$)
and temperature ($100{\rm K}<T<460{\rm K}$) range are described by the one band
intermediate coupling Hubbard model. The symmetrizarion method 
\cite{sym} describing the short range AF state $T<T^{\ast }$, is applied to
calculate both the spectral weight and conductivity. Surprisingly linear
resistivity, $\rho =\rho _{0}+AT$, is obtained with $A$ comparing well with
experiments\cite{Taillefer,Tl}. Quantitative comparison of the conductivity
with experiments therefore goes beyond scaling arguments\cite{QCP}. The
transition at $T^{\ast }$ is comparable to that observed in ARPES\cite%
{Greven} or transport. The transport versus spectroscopic $T^{\ast }$
determination is discussed within a well defined framework.

The single band Hubbard model is defined by the Hamiltonian:
\begin{eqnarray}
H &=&\sum \nolimits_{\mathbf{k,}\alpha =\uparrow ,\downarrow }a_{\mathbf{k}%
}^{\alpha \dagger }\left( \varepsilon _{\mathbf{k}}-\mu \right) a_{\mathbf{k}%
}^{\alpha }+U\sum \nolimits_{\mathbf{i}}n_{\mathbf{i}}^{\upharpoonleft }n_{%
\mathbf{i}}^{\downarrow };  \label{HubbardH} \\
\varepsilon _{\mathbf{k}} &=&-2t\left( \cos k_{x}+\cos k_{y}\right)
-4t^{\prime }\cos k_{x}\cos k_{y}-2t^{\prime \prime }\left( \cos 2k_{x}+\cos
2k_{y}\right) \text{.}  \notag
\end{eqnarray}%
The lattice spacing $a$ is the unit of length. Hoppings up to the third nearest neighbor are included  with values 
$t=250$ ${\rm meV}$, $t^{\prime }=-0.16t$, $t^{\prime \prime }=0.09t$. The chemical
potential $\mu $ varies in a wide range and the on-site $U=2.5t$.

Due to strong fluctuations in 2D, true long-range order exists only for
discrete symmetry breaking. Since the model possesses a continuous $SU\left(
2\right) $ spin symmetry, AF correlations are always short range.
Nevertheless, a well-defined crossover temperature $T^{\ast }$ exists that
separates the short range AF from the paramagnetic phase. At least naively,
the symmetry is \textquotedblleft almost" broken in the sense that the
correlator typically decreases slowly. The gaussian covariant approximation is the simplest variational approach 
which may account for the (spurious) dynamical symmetry breaking. It is described in details for bosonic and
fermionic systems in ref.\cite{Wang17}. In a recent paper\cite{sym} we
proposed a \textquotedblleft symmetrization" method to study strongly
interacting electronic systems in the pseudogap phase. Symmetrization is
achieved by integration of a one or two body correlator over the almost
broken symmetry group. The method was tested on the benchmark models, the 1D
and 2D one band Hubbard models.

One starts with a solution of the gaussian equations,
paramagnetic or AF\cite{Hirsch}, ignoring the spiral ones\cite{Igoshev16}),
and corrects it perturbatively by adding the leading self energy correction.
The method was originally proposed \cite{Thouless, RoandLi} in the
context of bosonic theories and is adapted to the fermionic case in SMI. The
correction gives the main contribution to the scattering rate determining the
conductivity, see SMII. All the frequency summations are performed exactly,
so that no problematic analytic continuation is needed. Full analytic
expressions in both the paramagnetic and pseudogap phases are also given. Of
course the range of validity of the expansion is limited by the requirement
that a high order correction (the order is rigorously defined in ref.\cite%
{RoandLi}) around the gaussian approximation should be smaller than the
preceding ones. As an example we present in Fig.~sm8 the inverse
compressibility of the Hubbard model with $U=6t$ and $t^{\prime }=-0.2t$, $%
t^{\prime \prime }=0$ at sufficiently high temperature, $0.2t<T<8t$, for
doping $0.15<p<0.3$. It almost coincides with the recent MC simulation \cite%
{Huang19} in this range of parameters. For a much smaller coupling $U=2.5t$
the lower bound of the applicable temperature is expected to extend down to
a much lower value, see SMIII.

The conductivity (per $CuO$ layer) is calculated by replacing Greens'
function in the standard Lindhard formula (derived from the Kubo formula in
SMII):
\begin{equation}
\sigma =-\frac{\pi e^{2}}{\hbar N}\int_{\omega =-\infty }^{\infty
}f_{F}^{\prime }\left( \omega \right) \sum \nolimits_{\mathbf{k}}v\left(
\mathbf{k}\right) ^{2}A\left( \omega ,\mathbf{k}\right) ^{2}\text{.}
\label{sigma}
\end{equation}%
Here $f_{F}\left[ \varepsilon \right] =\left( 1+\exp \left[ \varepsilon /T%
\right] \right) ^{-1}$ is Fermi distribution and the spectral weight is $%
A\left( \omega ,\mathbf{k}\right) =-1/\pi $ $\mathrm{Im}\left[ G\left(
\omega ,\mathbf{k}\right) \right] $. The (unrenormalized) Fermi velocity is $%
v_{i}\left( \mathbf{k}\right) =\frac{1}{\hbar }\frac{\partial \varepsilon _{%
\mathbf{k}}}{\partial k_{i}}$. It is important to note that the factor,
\begin{equation}
v\left( \mathbf{k}\right) ^{2}=4\left \{ \left( t+2t^{\prime }\cos
k_{y}+4t^{\prime \prime }\cos k_{x}\right) ^{2}\sin ^{2}k_{x}+\left(
k_{x}\leftrightarrow k_{y}\right) \right \} \text{,}  \label{vs}
\end{equation}%
vanishes quadratically at the van Hove singularities Gamma and M, see
Fig. sm4. 
As a consequence the anti-node region, the $\Gamma $ and the$\ M$ regions practically do not
contribute to the conductivity. It is important \textit{not to make} the
simplification of replacing the $f_{F}^{\prime }\left[ \omega %
\right] $ factor in Eq.(\ref{sigma}) by a delta function.

We have compared our method with recent MC simulation\cite{Huang19} for $%
t^{\prime }=-0.2t$, $t^{\prime \prime }=0$ at relatively high coupling $U=6t$
with doping $p=0.15-0.3$ and temperatures $T=0.2t-8t$. This shows that the
postgaussian approximation is still valid for such a large coupling\cite{sym}. 
The resistivity curves are slightly shifted, see Fig. sm9, compared to the
MC simulation. Obviously the case is not realistic since the values of $%
T^{\ast }$ determined from nonlinearity of $\rho \left( T\right) $ are
thousands of Kelvin ($\sim t$ taken as $250$ ${\rm meV}$) for low doping.
Analytically demonstrated \cite{highT} linearity in the large $T>>t$ limit
is quite common and is not directly related to the strange metal that
appears at temperatures $100-450{\rm K}$ which is obviously much lower than the
hopping energy $t$.
\begin{figure}[tbp]
\begin{center}
\includegraphics[width=8cm]{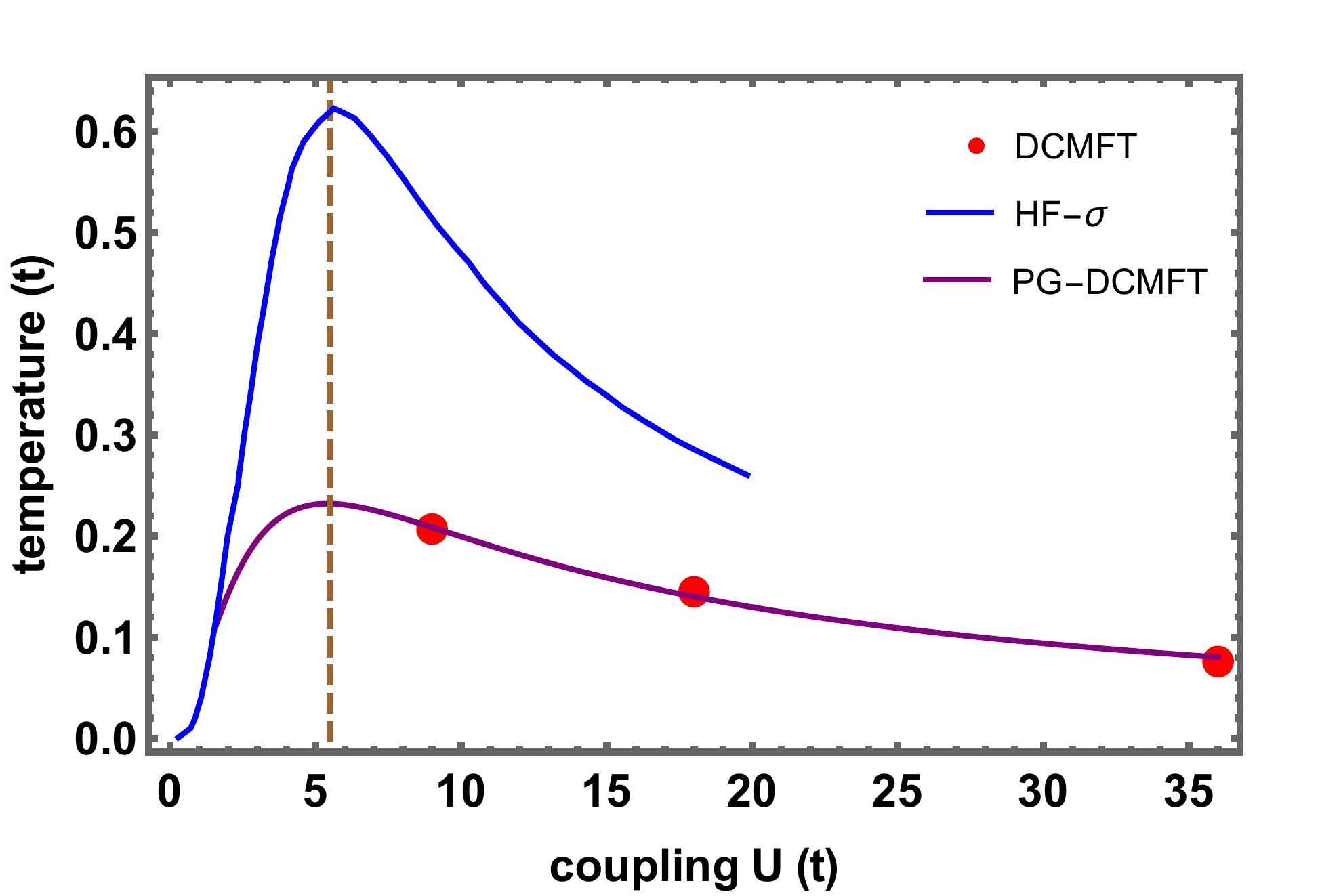}
\end{center}
\par
\vspace{-0.5cm}
\caption{The crossover temperature $T^{\ast }$ in the Hubbard model with ($%
t^{\prime }=0$) at half filling. The blue curve is the HF approximation at
small coupling and the nonlinear $\protect\sigma $ model at large coupling
interpolation. The purple curve is the gaussian perturbation theory at
intermediate coupling interpolated with the CDMFT diagrammatic approach at
strong coupling.}
\end{figure}

To characterize the pseudogap phase and the crossover line $T^{\ast },$
spectral weight, measured in numerous ARPES experiments\cite{Vishik18}, is
calculated. The spectral weight at the Fermi level as a function of quasi-momentum, $A\left( \omega =0,\mathbf{k}\right) $ for $p=0.21$ (calculated at
$220K$) is given in Fig.~2a. One notes that the spectral weight in the
antinodal region is larger than that in the nodal region by a factor of
about $2$.
\begin{figure}[tbp]
\begin{center}
\includegraphics[width=12cm]{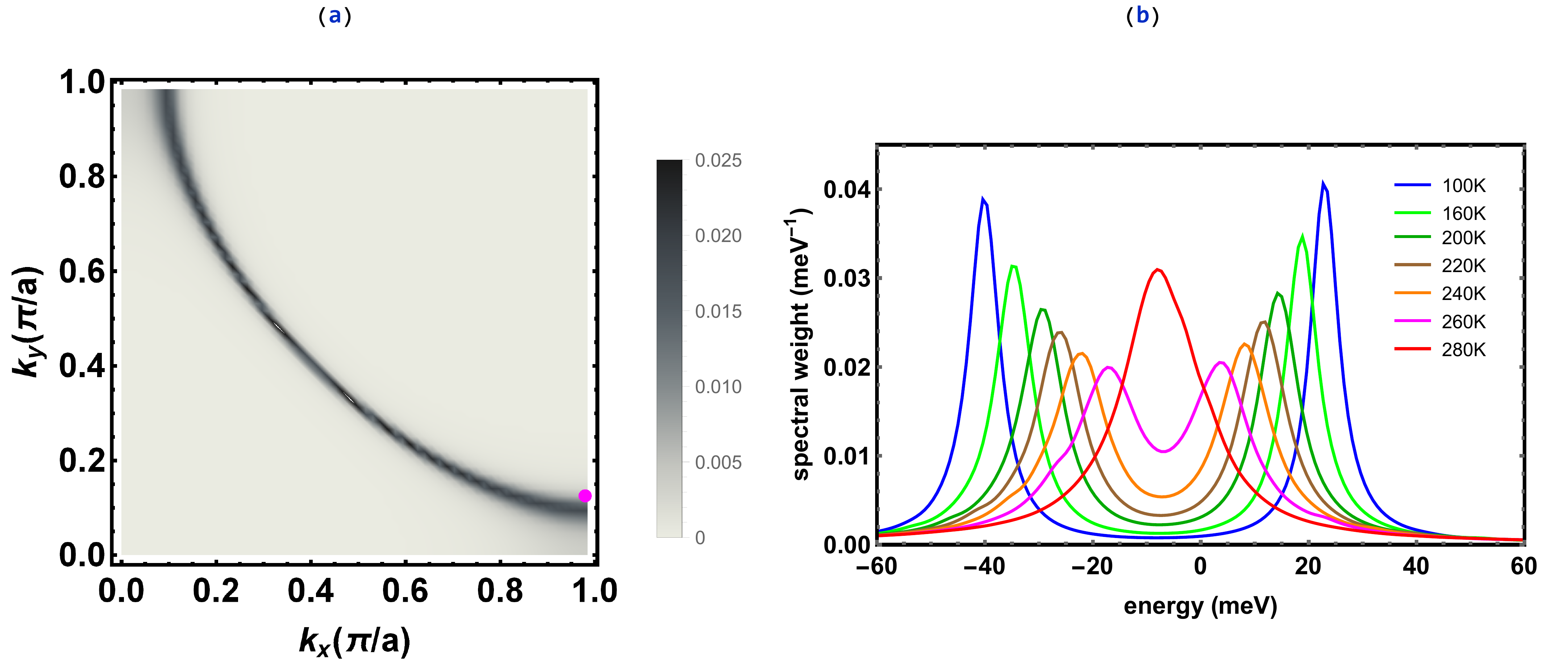}
\end{center}
\par
\vspace{-0.5cm}
\caption{(a) The zero frequency spectral weight in the first quarter of the
Brillouin zone. (b) Closing the pseudogap at a point close to the Fermi
surface of (a) (the pink blob) and near the anti-node. Spectral weight as a
function of frequency. The paramagnetic phase (a single peak) is represented
by $T=280{\rm K}$. The rest are anti-ferromagnetic (a double peak).}
\end{figure}

In Fig. 2b the spectral weight dependence on frequency across the $T^{\ast }$
crossover line for $p=0.16$ is given. We choose the quasi-momentum $\mathbf{%
k=}\pi \left( 1,\frac{1}{8}\right) $ close to the Fermi surface, see the
magenta blob in Fig.~2a. Several curves for the spectral weight with
temperature in the range $T=100-280{\rm K}$ are shown. The crossover $T^{\ast }$
therefore lies in the temperature range between $T=260{\rm K}$ and $280{\rm K}$. The
pseudogap value corresponds to the energy spacing between the two maxima. It
widens as the temperature decreases. Simultaneously the peak values become
higher and in the valley between them the spectral weight vanishes. These
features are qualitatively consistent with ARPES observations\cite{Vishik18}. 
The postgaussian correction modifies the HF picture by shifting $T^{\ast }$
to a lower value and reducing the pseudogap value at low temperature.
For other quasi-momenta (like near the nodal position given in Fig.~sm6) the
dependence is similar and in accord with available experiments.

Now let's turn to transport. The DC resistivity (per $CuO$ layer) is
given in Fig.~3. It clearly demonstrates the linear dependence, $\rho =\rho
_{0}+AT$, in the strange metal region of the phase diagram. This is the main
result of the present paper. The value of $A\simeq 25$ $\Omega /{\rm K}$
(resistance per layer) at doping $p=0.16$ are a bit higher than those found
by interpolating the $La_{2-p}Sr_{p}CuO_{4}$ data of \cite{Taillefer} to $%
p=0.16$. However, as was discussed in the Introduction,  the intermediate
value of $U=2.5t$ is smaller than that of $La_{2-p}Sr_{p}CuO_{4}$. We found
that for higher value of $U$, the slope decreases.  Since disorder is
always present in cuprates, doping-dependent value of disorder
strength $\eta _{0}=\hbar /2\tau _{0}$ must be taken into account. We have
chosen a small value of $\eta _{0}=3{\rm meV}$. Using a different value
of $\eta _{0}$ (see Fig.~sm5) would increase the residual resistivity $\rho
_{0}$. At higher doping the crossover to Landau liquid appears, see Fig.~3b.
\begin{figure}[tbp]
\begin{center}
\includegraphics[width=12cm]{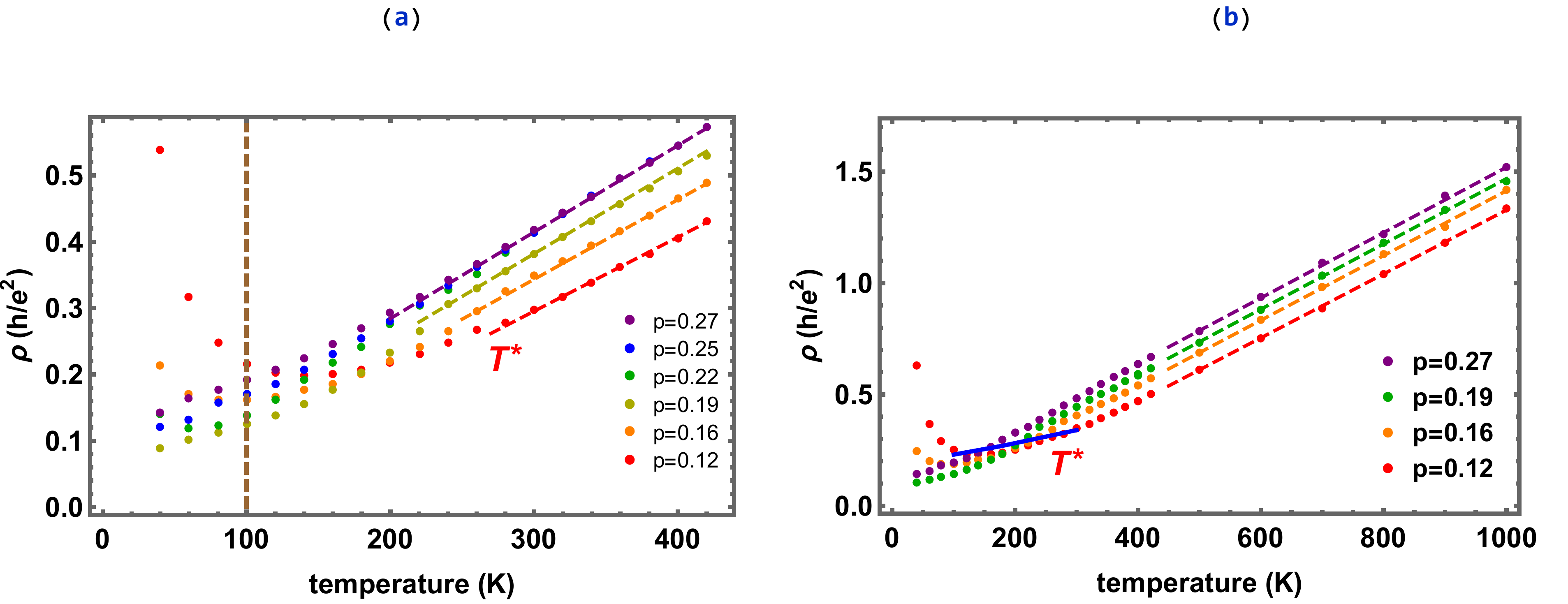}
\end{center}
\par
\vspace{-0.5cm}
\caption{Resistivity in the strange metal phase. (a) The postgaussian
results are very accurate beyond $100{\rm K}$ (dashed-brown line). Dashed straight
lines are interpolation. The transport determined crossover temperature is
marked by $T^{\ast }$. (b) A larger temperature and doping range demonstrate
crossover to the Landau liquid.}
\end{figure}

To understand qualitatively the results, we plot the integrand of the
imaginary part of the self energy (due to on site repulsion), $\mathrm{Im}%
\left( \Sigma _{k}\right) =\hbar /2\tau _{k}$, at $\omega =0$ as a function
of quasi-momentum in Fig.~4a,b for two different temperatures. For this
purpose a higher doping $p=0.26$ is considered and thus there would be no
"intercept" $\rho _{0}$ in Fig.~3. The lower doping case will be discussed
in SMII. Plots of the ratio $A\left( \mathbf{k}\right) /T$ over the whole BZ
for $T=200{\rm K}$, $400{\rm K}$ are shown in Fig.~4 and the curves are hardly
distinguishable. This demonstrate that $1/2\tau _{k}\propto T$. Due to the
maximum of the factor $v\left( \mathbf{k}\right) ^{2}$ in Eq.~(\ref{sigma})
most important contribution to conductivity comes from a broad region near
the nodal point. To demonstrate this, we plot the spectral weight dependence
along the $\Gamma -M$ line (see the pink line in Fig.~4a) in Fig.~4c. The
dependence on location is different for temperatures $200{\rm K}, 300{\rm K}, 400{\rm K}.$
However, all the three curves almost coincide, when the spectral weight is
divided by $T$, as shown in Fig.~4d. One observes from Fig.~2a that the
Fermi surface in this case is nearly circular. Thus the effective mass
approximation can be applied to estimate the conductivity via the Drude
formula $\sigma =e^{2}n\tau /m^{\ast }$ with $\hbar /2\tau =\mathrm{Im}\left[
\Sigma _{node}\right] \simeq 0.25T$. This is close to the result in Fig.~3.
Using the phenomenological (Planck) formula $A=\alpha m^{\ast }/e^{2}n\hbar $
\cite{Taillefer}, one estimates $\alpha =0.5$.

\begin{figure}[hbt!]
\includegraphics[width=12cm]{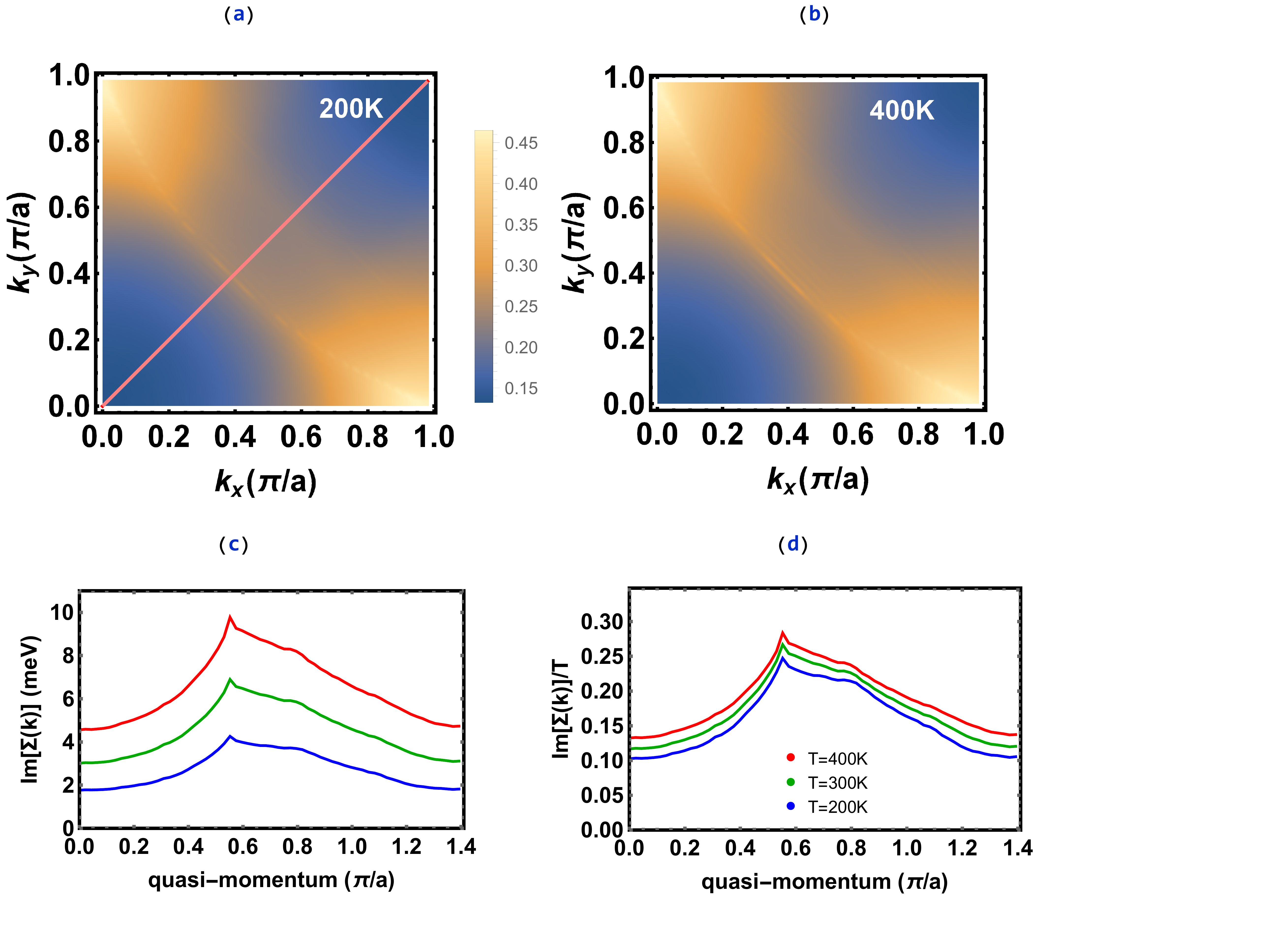}
\caption{(a) The self energy due to Hubbard repulsion for doping $p=0.26$.
The imaginary part of the self energy divided by temperature, $\mathrm{Im} %
\left[ \Sigma \left( 0,\text{k}\right)/T \right] $, in the first quarter of
the BZ for 200{\rm K}. (b) Same for 400{\rm K}. Note the maximum on the Fermi surface
(shown in Fig.~2a). (c) The imaginary part of the self energy for
temperatures at 200{\rm K}, 300{\rm K}, and 400{\rm K} on the $\Gamma$ to $M$ line (marked
by a pink line in (a)). (d) The imaginary part of the self energy divided by
temperature for the same temperature.}
\label{fig4}
\end{figure}

To summarize, an intermediate coupling one band Hubbard model can be used to
describe the two most important unusual normal state features of cuprates: the pseudogap and strange metal. Both the
spectroscopic and transport properties of the cuprates in the whole doping
range were considered on the same footing within a relatively simple
(symmetrized) postgaussian approximation. It is valid for intermediate
couplings $U/t=1-4$ in the temperature range $T=100-500{\rm K}$. We have
assumed a relatively small coupling that is independent of doping in
the range $0.1-0.3$. For a smaller doping, especially in the Mott insulator
phase, the coupling on the effective tight binding scale increases and a
different method would be required. This provides an alternative to the
commonly accepted paradigm that the coupling at a significant doping should
be strong enough, say $U/t>6$, for the system to describe the strange metal.
We argued (see also description of the Lifshitz transition and
fractionalization of the Fermi surface in a similar framework\cite{phonon})
that to obtain phenomenologically acceptable underdoped normal state
characteristics like $T^{\ast }$, pseudogap values, and spectral weight
distribution, a large value of $U$ is detrimental. Of course this applies
only to the short range AF order interpretation of the
pseudogap. Surprisingly the resistivity in the above temperature range is
linear $\rho =\rho _{0}+\alpha \frac{m^{\ast }}{e^{2}n\hbar }T$ with the
"Planckian" coefficient $\alpha \sim 0.5$. Interestingly the spectroscopy
estimate of $T^{\ast }$ (from the vanishing of the pseudogap, Fig.~2b, as
observed in ARPES) is typically lower than that determined from resistivity
(Fig.~3).

First principle calculations for parent materials of the electron doped
cuprates generally result in intermediate or even small values
of the effective $U$\cite{Jang16}. The present study demonstrates that for hole doped
cuprates the intermediate coupling option is viable despite the fact that
most first principle determinations for \textit{parent materials } favour a
large coupling \cite{Nilsson19}. Materials like $La_{2}CuO_{4}$, $%
Bi_{2}Sr_{2}CuO_{6}$ perhaps are really strongly coupled even when doped,
but higher $T_{c}$ superconductors like $HgBa_{2}CuO_{4}$, $%
Tl_{2}Ba_{2}CuO_{6}$ might belong to the intermediate coupling class when
doped. The situation with two or three layered cuprates should be similar to the
model adapted to include inter-layer hopping.

The authors are very grateful to J. Wang, B. Shapiro, Z. Sun for numerous
discussions. We would also like to thank the National Center for
High-performance Computing (NCHC) of National Applied Research Laboratories
(NARLabs) in Taiwan for providing computational and storage resources. H.C.K
and B. R. are supported by MOST-110-2112-M-A49 -012 -MY2 of MOST, Taiwan. D.
P. L. was supported by National Natural Science Foundation of China (No.
11674007 and No. 91736208).

\end{document}